\documentclass[intlimits,twoside,a4paper]{article}

\usepackage[T2A]{fontenc} 
\usepackage[utf8]{inputenc} 
\usepackage[greek,ukrainian,english]{babel}
\usepackage{cmpj3}

\usepackage{amsmath}
\usepackage{empheq}
\usepackage{pstricks}
\usepackage{mathrsfs}
\usepackage{mathtools}
\usepackage{cancel}
\usepackage{accents}
\usepackage{dsfont}
\usepackage{color}
\usepackage[normalem]{ulem}
\usepackage[caption = false]{subfig}

\usepackage{upgreek}

 \DeclareRobustCommand{\augiefamily}{%
 	\fontfamily{augie}\fontseries{m}\fontshape{n}\selectfont}
 \DeclareTextFontCommand{\textaugie}{\augiefamily}

\def\be{\begin{equation}}
	\def\ee{\end{equation}}
\def\bea{\begin{eqnarray}}
	\def\eea{\end{eqnarray}}
\def\bpar{\left(\!\!\begin{array}}
	\def\epar{\end{array}\!\!\right)}
\def\bdar{\left|\!\!\begin{array}}
	\def\edar{\end{array}\!\!\right|}
\def\barr{\begin{array}}
	\def\earr{\end{array}}
\def\btab{\begin{tabular}}
	\def\etab{\end{tabular}}

\def\<{\langle}
\def\>{\rangle}

\issue{2024}{27}{3}{33801}
\doinumber{10.5488/CMP.27.33801}

\title[Consensus decision making on a complete graph]{Consensus decision making on a complete graph: complex behaviour from simple assumptions}

\author[P. Sarkanych, Yu. Sevinchan, M. Krasnytska, P. Romanczuk, Yu. Holovatch]{
			{P. Sarkanych}\orcid{0000-0001-6916-5004} \refaddr{inst1,inst2}, 
			{Yu. Sevinchan}\orcid{0000-0003-3858-0904}\refaddr{inst3,inst4}, 
			{M. Krasnytska}\orcid{0000-0002-0464-5741}\refaddr{inst1,inst2,inst7}, 
			{P. Romanczuk}\orcid{0000-0002-4733-998X}\refaddr{inst3,inst4}, 
			{Yu.~Holovatch}\orcid{0000-0002-1125-2532}\refaddr{inst1,inst2,inst5,inst6}
    }
\addresses{
		\addr{inst1}{{Institute for Condensed Matter Physics of the National Academy of Sciences of Ukraine}, 
			{Lviv},
			{79011}, 
			{Ukraine}}
   \addr{inst2}{{$\mathbb{L}^4$ Collaboration and Doctoral College for the Statistical Physics of Complex Systems}, 
			{Lviv-Leipzig-Lorraine-Coventry}, 
			{Europe}}
   \addr{inst3}{{Department of Biology, Institute for Theoretical Biology, Humboldt Universit\"at zu Berlin}, 
			{Berlin}, 
             {10099},
			{Germany}}
   \addr{inst4}{{Research Cluster of Excellence `Science of Intelligence'}, 
			{Berlin}, 
             {10587}, 
			{Germany}}
		\addr{inst7}{Haiqu, Inc., Shevchenka Str., 120 G, Lviv, 79039, Ukraine}
		\addr{inst5}{{Centre for Fluid and Complex Systems, Coventry University}, 
			Coventry CV1 5FB,
			{UK}}
		\addr{inst6}{Complexity Science Hub Vienna, 1080 Vienna, Austria} 
		}

\date{Received February 05, 2024, in final form March 15, 2024}

\begin{document}
\maketitle
\begin{abstract}
				
   In this paper we investigate a model of consensus decision making [Hartnett~A.~T., et al., {Phys. Rev. Lett.}, 2016, \textbf{116}, 038701] following a statistical physics approach presented in [Sar\-ka\-nych P., et al., {Phys. Biol.}, 2023, \textbf{20}, 045005].
   Within this approach, the temperature serves as a measure of fluctuations, not considered before in the original model.
   Here, we discuss the model on a complete graph. The main
   goal of this paper is to show that an analytical description 
   may lead to a very rich phase behaviour, which is usually not expected for a complete graph.
  However, the variety of individual agent (spin) features --- their inhomogeneity and bias strength --- taken into account by the model leads to rather non-trivial collective effects.
  We show that the latter may emerge in a form of continuous or abrupt phase transitions 
  sometimes accompanied by re-entrant and order-parameter flipping 
  behaviour. In turn, this may lead to appealing interpretations in terms 
  of social decision making. We support analytical predictions by numerical 
  simulation. Moreover, while analytical calculations are performed within an
  equilibrium statistical physics formalism, the numerical simulations 
  add yet another dynamical feature --- local non-linearity or conformity of the individual 
  to the opinion of its surroundings. This feature appears to have a strong
  impact both on the way in which an equilibrium state is approached as well as
  on its characteristics. 
		\end{abstract}

\hspace{20em}\textit{{In memory of Ralph Kenna: wonderful person,}} 

\hspace{20em}\textit{{brilliant physicist, and very dear friend.}}

\section{Introduction}	\label{I}	


Collective behaviour is a fascinating phenomenon exhibited by a wide range of biological taxa, ranging from social insects \cite{bonabeau1997self,buhl2006disorder} to humans crowds \cite{faria2010collective}. Different evolutionary benefits of being in a group have been discussed in the literature, such as increased predator protection or improved foraging success \cite{krause2002living,sumpter2010collective}. Many of these benefits can be traced back to collective information processing and collective decision making, i.e., the ability of animal collectives to process and pool information gathered by individuals through social interactions, and to make accurate consensus decisions in a fully distributed fashion \cite{couzin2005effective,couzin2011uninformed,winklmayr2020wisdom,winklmayr2023collective}. The corresponding remarkable capability of collectives to outperform even the best individuals within the group led to the introduction of terms ``collective intelligence'' or ``wisdom of crowds''.   

From a statistical physics point of view, the process of an initially undecided group choosing one option among multiple available ones (typically two), corresponds to a symmetry-breaking phase transition \cite{romanczuk2023phase}. Therefore, spin models provide a natural and well established theoretical framework to study the self-organised collective decision making in systems of many interacting agents \cite{castellano2009statistical,pinkoviezky2018collective}.
In 2016, Hartnett~{et al.}~\cite{Hartnett16} suggested a spin model for binary collective decision making in heterogeneous collectives, consisting of subgroups with different biases towards one of the two options. The modelling work was inspired by previous experimental and theoretical results on the decisive role of unbiased individuals in the system as ``mediators'' between subgroups with opposing biases.
In the original publication, the model was studied on the lattice, which demonstrated a rich phenomenology. Recently, we have shown that the consideration of the model on a complete graph enables an analytical investigation of the consensus states in the thermodynamic equilibrium \cite{Sarkanych23}. Here, we build on this previous work, and demonstrate how a systematic analysis of the steady-state dynamics exhibits a rich behaviour, even on the complete graph. Our particular focus lies on the role of temperature, not considered when the original model was introduced.

In the next section, we provide a more in-depth discussion of the model, before turning to the theoretical predictions and to the results of numerical simulations.

\section{The model: agent-individual features and local decision making rules} \label{II}


The model suggested by Hartnett~{et~al.}~\cite{Hartnett16} to describe consensus decision making in a group of agents rests on two main ingredients, which are meant to account for agent heterogeneity and for the non-linearity governing their local interaction dynamics. Below we explain them more in detail.

\subsection{Heterogeneity: a bias in the preferred state}\label{IIa}

\begin{figure}[htbp]
    \includegraphics[angle=0,origin=c,width=0.45\linewidth]{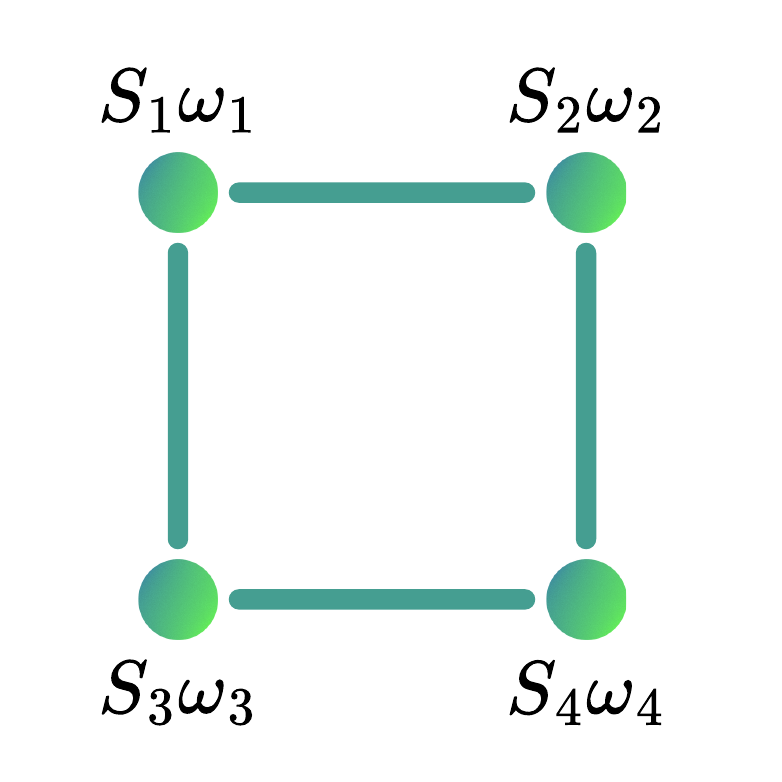}
    \includegraphics[angle=0,origin=c,width=0.45\linewidth]{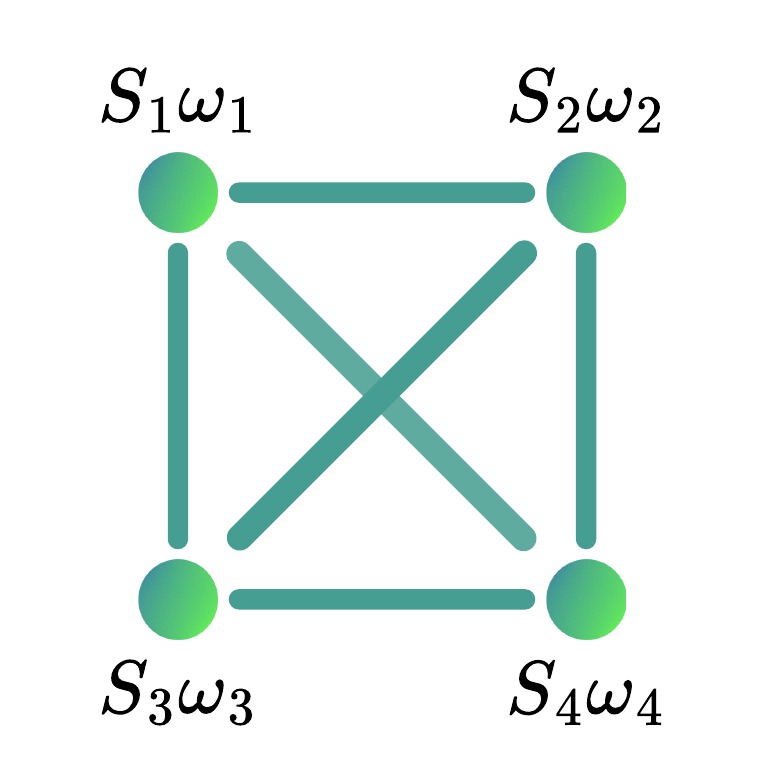}
    \centerline{\textbf{(a)} \hspace{22em}   \textbf{(b)}}
    \caption{(Colour online) 
        In the model we consider, each of $N$ agents $S_1, \dots , S_N$ is able 
        to attain one of the two contradicting opinions: $S_i=\pm 1$ ($N=4$ in the figure). 
        Besides, each agent~$i$ can have a bias~$\omega_i$ towards any opinion 
        or can  be unbiased: $\omega_i=\{\omega_+,\omega_-,\omega_0 \}$.
        Agent biases and initial opinions are randomly distributed and fixed in a given configuration.
        The biases remain unchanged, these are individual opinions that change when the consensus is reached under the influence of a social field, equation~(\ref{2.1}), experienced by each agent.
        \textbf{(a)} Agents are located on a 2D square lattice, only the nearest neighbours emanate social field for a given agent (the case originally suggested in reference~\cite{Hartnett16}). 
        \textbf{(b)}~Each agent experiences social field emanated by all other agents (the case analysed in reference~\cite{Sarkanych23} and in this paper). 
    }
    \label{fig1}
\end{figure}

Let us consider a process of reaching consensus in a group of $N$~agents, each being able to attain two opinions (states) $S_i=\pm 1$, $i=1,\dots,N$. To account for agent heterogeneity, let us attribute an additional feature to each of them, reflecting agent bias in the preferred state and representing the strength of an individual opinion.
Below, we call the biased agents the `informed' ones, assuming that their bias originates from the information about a preferred opinion/state.
In turn, unbiased/uninformed agents do not have any individual preference. 
This feature is described by a three-state random variable $\omega_i=\{\omega_0,\omega_+,\omega_-\}$.
The value $\omega_0=1$ corresponds to an unbiased agent (no preferred states), whereas $\omega_+>1$ and $0<\omega_-<1$ correspond to agent preferences/biases to the $S_i=+1$ and $S_i=-1$ opinions, respectively. 
Note that for the negatively biased agents, a smaller $\omega_-$ corresponds to a stronger bias towards the opinion $S_i=-1$.
The individual biases are randomly and uniformly distributed with densities $\rho_0, \, (1 - \rho_0)\rho_+,\, (1 - \rho_0)\rho_-$, where $\rho_++\rho_-=1$ are partial densities of biased agents and $\rho_0\leqslant 1$ is the density of unbiased/uninformed ones.
See also figure~\ref{fig1}. 

Agents interact via the local social field~$h_i$ that acts on each of them and is
defined based on the opinions of the neighbourhood:

\begin{equation}
    \label{2.1}
    h_i=\frac{\omega_i n_i^+ - n_i^-}{\omega_i n_i^+ + n_i^-}.
\end{equation}
Here, $n_i^\pm$ are numbers of the $i$-th agent nearest neighbours with the opinion $+1$ or $-1$, correspondingly.

\subsection{Local non-linearity: conformity to the neighbouring opinion}\label{II.b}

\begin{figure}[htbp]
    \centerline{
        \includegraphics[width=0.4\linewidth]{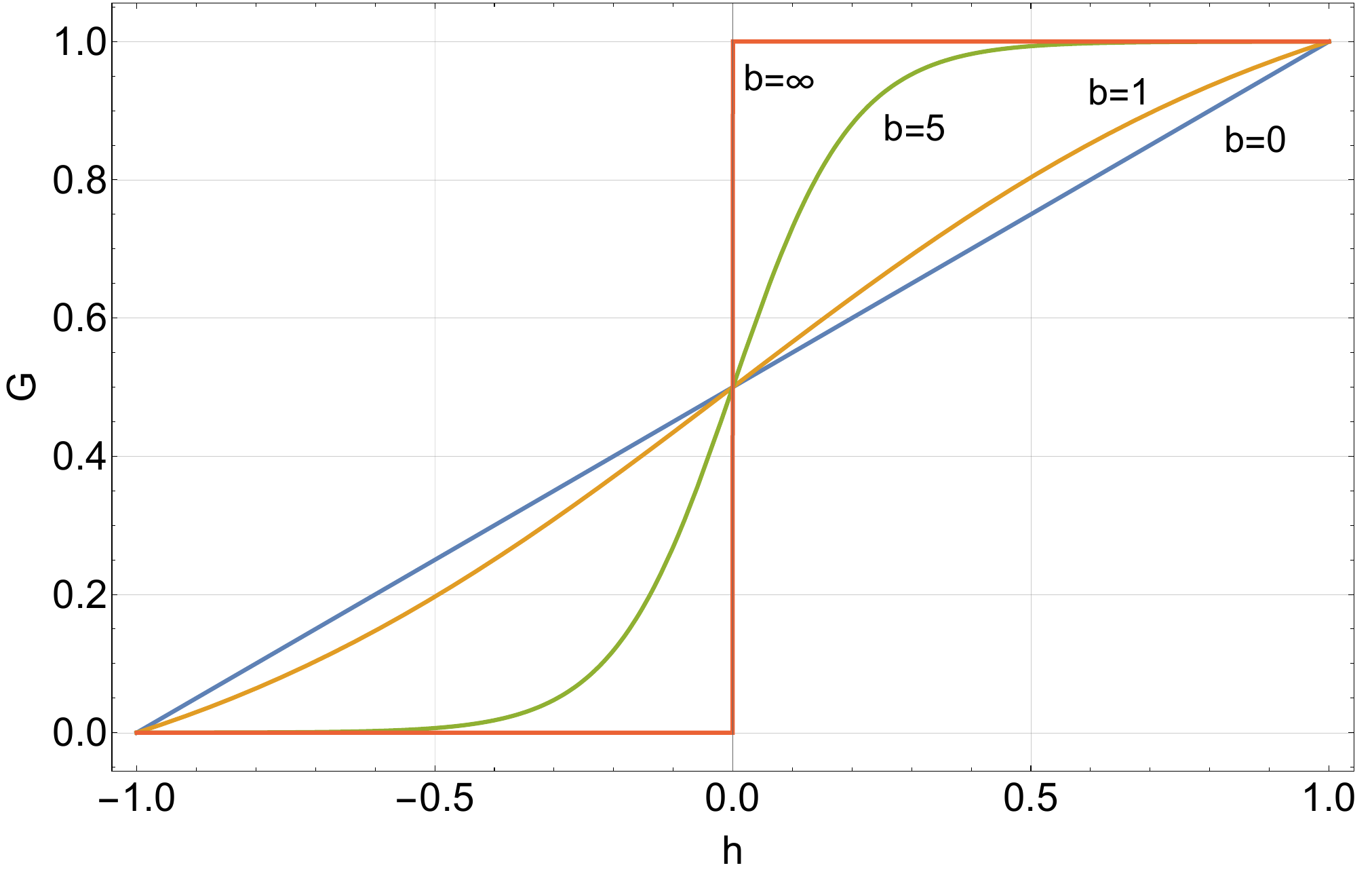}\hspace{3em}
        \includegraphics[width=0.4\linewidth]{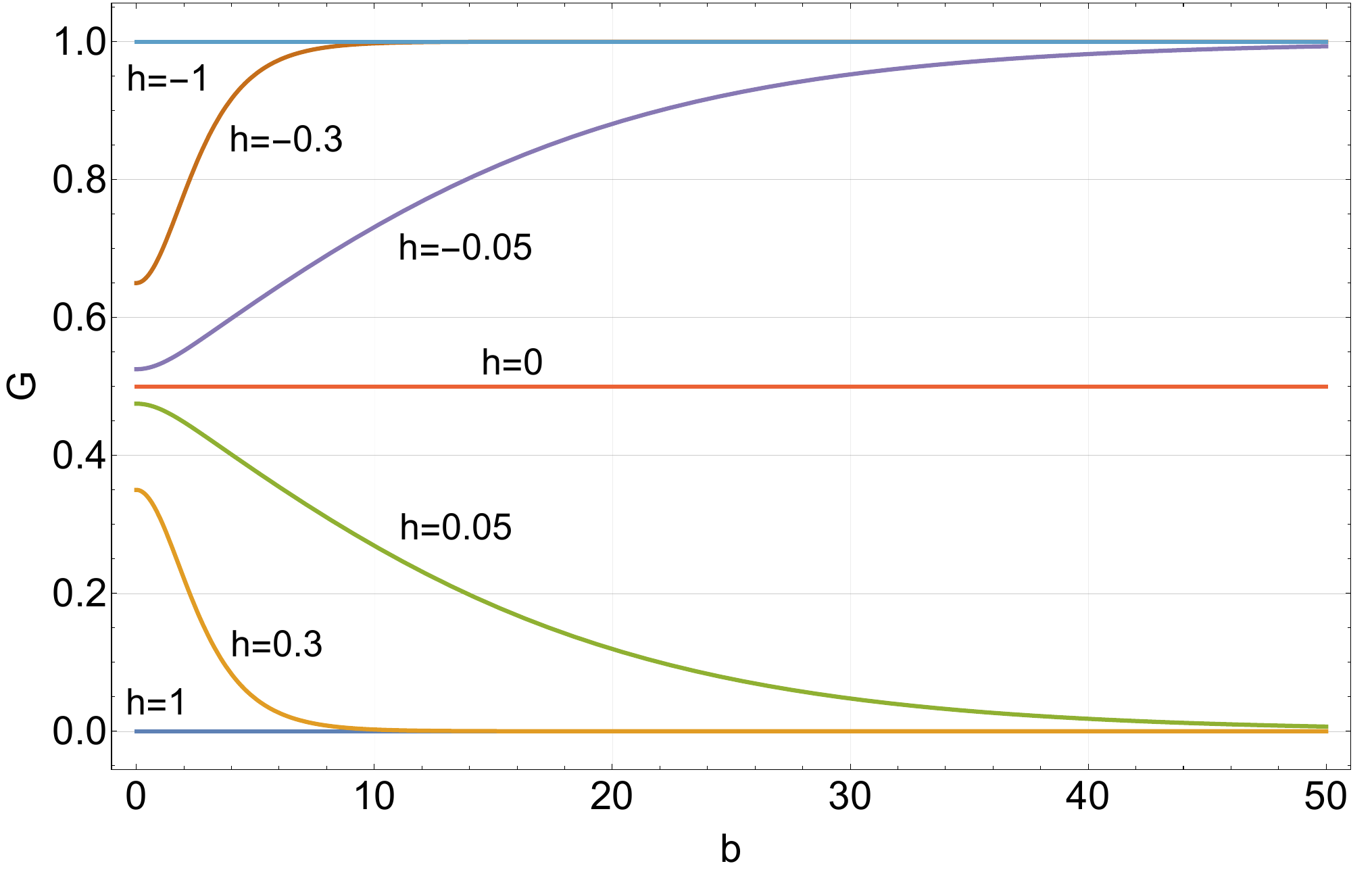}
    }
    \centerline{\textbf{(a)} \hspace{15em}  \textbf{(b)}}
    \caption{(Colour online) 
        Behaviour of the probability function $G$~[equation~(\ref{2.2})]
        for different values of the social field $h$ and of the non-linearity parameter $b$.
        Note that by definition [equation~(\ref{2.1})] the social field is bounded in the region $-1\leqslant h \leqslant 1$.
        \textbf{(a)}. At small $b$, the dependency on~$h$ is almost linear, this changes
        to the sigmoid form with an increase of $b$, finally resulting in a step-like
        function for large~$b$. 
       \textbf{(b).} At large $|h|\sim 1$, the function does not depend on $b$: $G_i \sim 1$, $G_i \sim 0$ for $h\sim \pm 1$, correspondingly. With a decrease of $|h|$, $G$ strongly discriminates between large and small $b$, whereas for small $|h|\sim 0$ it attains a monotonous behaviour reaching $G=1/2$ at $h=0$.
    }
    \label{fig2}
\end{figure}

Being $\omega$-dependent, the social field~(\ref{2.1}) is distorted by an individual bias: it is enhanced when the neighbours are of the same opinion as the agent bias and it is weakened when these opinions do not coincide.
Such an effect is known as a confirmation bias, widely discussed in the literature \cite{klayman1995varieties,nickerson1998confirmation}.
However, in general, an individual can respond to a given social field in different ways.
Possible diversities of such a response constitute another ingredient of the model. 
To this end, to interpret the model evolution as consensus decisions making,
an algorithm has been proposed that probabilistically defines the state of an agent~$i$ at time~$t+1$ by its social field $h_i$ at time $t$ with the
help of the probability function
\begin{equation}
    \label{2.2}
    G_i=\frac{1}{2}\bigg[ 1 + \frac{\tanh (b h_i)}{\tanh (b)}\bigg]\, ,
\end{equation}
with the nonlinearity parameter~$b$.
Within the consensus making dynamics, agents in state $S=-1$ switch to the state $S=+1$ with a probability $G_i$ and agents in state $S=+1$ switch to the state 
$S=-1$ with a probability $1-G_i$.
When the bias is absent (all $\omega_i=1)$, equation~(\ref{2.2}) in the limiting cases $b=0$ and $b=\infty$ reproduces dynamics rules of the classical voter \cite{Clifford73,Redner19} and majority-rule \cite{Krapivsky03} models, correspondingly.
Its behaviour is further sketched in figure~\ref{fig2}.
Since function~$G_i$ describes agents' reaction on the opinions/states of its neighbours, it may be also interpreted in terms of agent conformity, i.e., its tendency to align with the opinions of its surroundings.
As one sees from figure~\ref{fig2}, such a reaction is governed by parameter~$b$ in a non-trivial way, thus allowing to treat $b$ as a non-linearity parameter, further discussed below.

Taking into account the agent non-homogeneity and local non-linearity allowed to unveil
fundamental mechanisms that govern the decision making dynamics also in living systems.
In particular, it has been shown~\cite{Hartnett16} that uninformed (unbiased)
agents may play a key role in the decision making process while the final collective
state is strongly dependent on the non-linearity of local interactions.
Originally, the model was described in terms of a generalised voter model, by suggesting a cellular automata algorithm to run its evolution. In our recent paper 
we pushed forward an analogy between the decision making process and the phase transition 
into the ordered state and suggested a statistical physics description, writing down the 
many-particle Hamiltonian governing the inter-agent interactions~\cite{Sarkanych23}.
In terms of such a description, temperature serves as a measure of fluctuations, not considered before in the original model.
It appeared that different Hamiltonians that capture principal features of the above described algorithm can be suggested.
In what follows below we use one of such Hamiltonian descriptions to show that the interplay of agent bias and conformity --- their inhomogeneity and non-linearity --- taken into account by the model leads to rather non-trivial collective effects that were not observed so far.

\section{Complex behaviour from simple assumptions}\label{III}

\subsection{Equilibrium state. Analytic results for the phase diagram}\label{III.1}

In reference~\cite{Sarkanych23} we have complemented an analysis of the algorithmic model described above 
for opinion formation by using the approach of statistical physics.
Doing so, on the one hand, we had the intention to push the physical analogy of the process of opinion formation even further; on the other hand, we had in mind to open an avenue to explore the influence of possible noise/random fluctuations in the agent behaviour on the entire process.
This last aim can be achieved by interpreting a temperature variable as a source of such noise.
One of the Hamiltonians that is suggested in reference~\cite{Sarkanych23} (and called the social-field Hamiltonian there) reads:
\begin{equation}\label{3.1}
    H =-\sum_{i=1}^{N} h_iS_i\, ,
\end{equation}
where the local field $h_i$ is given by equation~(\ref{2.1}) and $S_i=\pm 1$, as before. 
In this Hamiltonian, the interaction between spins~$S_i$ occurs via the field~$h_i$ 
that is emanated by its neighbours.
Note, however, that the `social field'~(\ref{2.1})  is introduced in the algorithm by the update rule~(\ref{2.2}). The last, in turn, involves
the conformity parameter~$b$ which is of course absent in the equilibrium statistical 
physics settings. We will comment more on that later.

Considering the model Hamiltonian~(\ref{3.1}) on the complete graph when each agent is a subject of a social field emanated by all other agents (see figure~\ref{fig1} \textbf{b}), we used the mean-field approach to obtain expressions for the thermodynamic functions in a closed form.
In particular, the obtained Gibbs free energy per site reads:
\begin{equation} \label{3.2}
    - \beta g(\beta)= \rho_0\log[\cosh(\beta m )] + 
    (1-\rho_0)\rho_+\log[\cosh(\beta h_+)] + 
    (1-\rho_0)\rho_-\log[\cosh(\beta h_-)] 
     ,
\end{equation}
where $\beta=1/(k_{\rm B} T)$ is an inverse temperature, $k_{\rm B}$ is the Boltzmann constant, and
the equilibrium magnetisation per site (consensus opinion) $m=m(\beta)$ is found from 
the solution of the self-consistency relation:
\begin{equation}\label{3.3}
    m(\beta) = \rho_0\tanh(\beta m) +  (1-\rho_0)\rho_+\tanh(\beta h_+ ) + 
    (1-\rho_0)\rho_-\tanh(\beta h_-)
    ,
\end{equation}
with fields
\begin{equation}
    h_\pm = \dfrac{(\omega_\pm + 1)m+\omega_\pm-1}{(\omega_\pm - 1)m+\omega_\pm+1} .    
\end{equation}

In what follows below, we aim to show that the thermodynamics of a model governed
by the Hamiltonian~(\ref{2.1}) is characterised by a rich phase behaviour, which is usually not expected for models on a complete graph. Such a behaviour is due to the variety of individual agent (spin) features --- in this case, it is their inhomogeneity caused by
bias strength --- taken into account by the model.
To demonstrate this, we will have a closer look on numerical solutions of a transcendental equation~(\ref{3.3}) 
for several choices of model parameters ($\rho_i, \omega_i$). These case studies may be further interpreted as: 
two {weakly} biased groups of similar size; strongly biased minority and weaker majority;
confronting strongly biased groups.
The main difference between these three cases is the ratio between the bias strength $\omega_+$ and $\omega_-$, 
as well as the composition of the system governed by densities $(\rho_0,\rho_+, \rho_-)$.\footnote{Note that $\omega_+$ and $\omega_-$ are used on inverse scales, thus in order to understand which subgroup has stronger bias, one has to compare $\omega_+$ and $1/\omega_-$. 
In addition only two out of three densities are independent as $\rho_++\rho_-=1$.}
We will have a closer look on the equilibrium properties of each of these three special cases in the following subsections.

\subsubsection{Competition between two weakly biased groups of similar size}

\begin{figure}[htbp]
	\centerline{
		\includegraphics[width=0.45\textwidth]{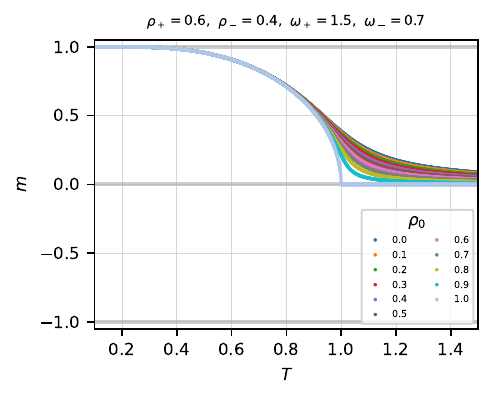}
		\includegraphics[width=0.45\textwidth]{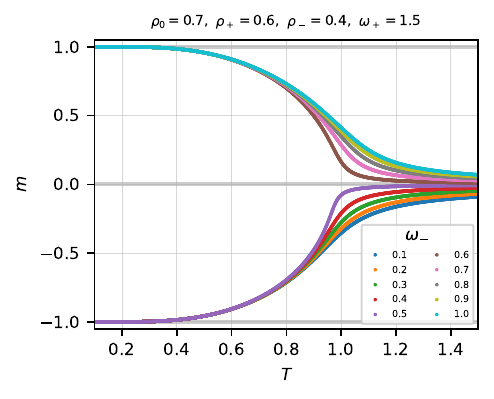}
	}
	\centerline{\textbf{(a)} \hspace{15em} \textbf{(b)}}
	\caption{(Colour online) Equilibrium magnetisation as a function of temperature in the case when the system consists of {weakly} biased groups comparable by size. The behaviour is identical to the Ising model in an external magnetic field. 
		\textbf{(a)} The effect that the fraction of unbiased individuals has on the consensus opinion between two {weakly} biased groups comparable by size. As the fraction of unbiased individuals $\rho_0$ increases, the equilibrium magnetisation decreases. Only in the limiting case $\rho_0=1$ the system undergoes the second order phase transition. 
		\textbf{(b)} The effect of the bias of the smaller subgroup has on the consensus formation between two {weakly} biased groups comparable by size. When the smaller subgroup biased towards $S=-1$ has a stronger bias (low values of $\omega_-$) it manages to shift the consensus opinion to the negative values of magnetisation. Otherwise, the majority dictates the decision. It is similar to the change of a sign of the effective external magnetic field.
	}
	\label{fig3}
\end{figure}

We will illustrate this case by choosing $\rho_+=0.6$, $\rho_-=0.4$ and $\omega_+=1.5$. 
Here, we purposely allow $\rho_0$ and $\omega_-$ to change to examine how they affect the 
equilibrium properties. 
It was earlier shown~\cite{Hartnett16} that unbiased individuals help the system to 
reach the consensus.
Therefore, in figure~\ref{fig3}(a), we show how equilibrium magnetisation (consensus opinion) depends on the temperature for different values of $\rho_0=[0,1]$ and 
fixed $\omega_-=0.7$.
In the limiting case $\rho_0=1$, all the agents are unbiased, thus the system reverts to the standard Ising model. 
It has a twofold degeneracy of the ground state dictated by the symmetry of the $S=+1$ and $S=-1$ states. 
For this plot, we omit the negative branch for clarity reasons.

Based on the curves in figure~\ref{fig3}(a), one can tell that for small temperatures the difference in equilibrium magnetisation is very small and becomes more pronounced as temperature rises. 
Furthermore, the introduction of biased individuals destroys the phase transition that exists in the Ising limiting case making the system behave like the Ising model in an external magnetic field, i.e., there is always some finite magnetisation for any finite temperature. 
The higher the fraction of unbiased individuals, the smaller is the resulting magnetisation for a fixed temperature.

In figure~\ref{fig3}(b) we examine the effect of the bias on the consensus opinion. 
To this end, we fix $\rho_0=0.7$ and plot magnetisation as a function of temperature for $\omega_-=[0.1,1]$.
In the case of strong bias (low values of $\omega_-$), the consensus opinion is aligned with the $S=-1$ state, while for weaker biases the majority $\rho_+>\rho_-$ favouring $S=+1$ wins. 
Similarly to figure~\ref{fig3}(a) the system behaves like the Ising model in the external field, {with a small exception that in a narrow range $0.54\leqslant\omega_-\leqslant0.55$ a first order transition occurs. This effect will be addressed in the next subsection.} 

\subsubsection{Competition between strongly biased minority and weaker majority}

For this case, we will assume that our agents have stronger biases than in the previous case and their densities are comparable but not equal. 
Obviously, if the dominating subgroup has a stronger bias, it dictates the consensus opinion.
Therefore, we will consider the case where the larger subgroup has a slightly weaker bias than the smaller subgroup. 
In the following example we consider $\rho_+=0.6$, $\rho_-=0.4$, $\omega_+=2$, varying parameters $\rho_0$ and $\omega_-$.

\begin{figure}[htbp]
	\centerline{
		\includegraphics[width=0.45\textwidth]{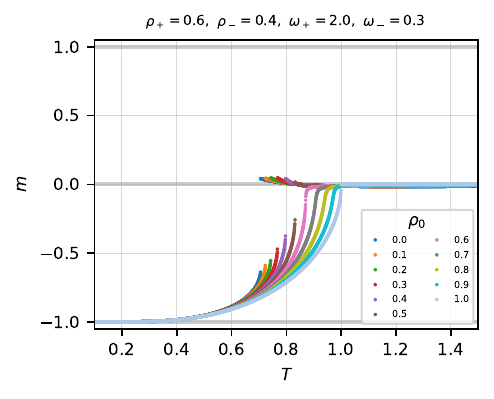}
		\includegraphics[width=0.45\textwidth]{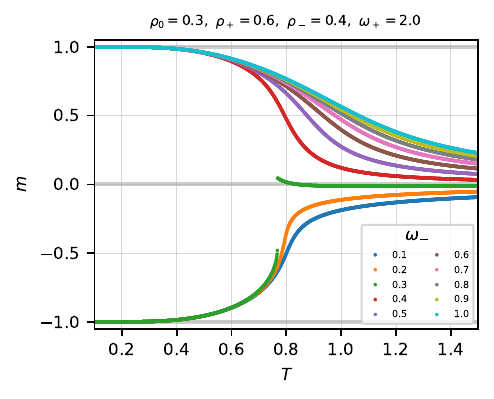}
	}
	\centerline{\textbf{(a)} \hspace{15em} \textbf{(b)}}
	\caption{(Colour online) Equilibrium magnetisation as a function of temperature in the case when the system consists of a larger subgroup with the weaker bias and a smaller subgroup with a stronger bias. In this case, the system may undergo a first order phase transition. 
		\textbf{(a)} The effect that the fraction of unbiased individuals has on the consensus opinion within strongly biased minority and weaker majority. Similarly to the previous case shown in figure~\ref{fig3}(a), an increase in the number of unbiased individuals results in a more aligned equilibrium state for the temperature range outside transition region. For low values of $\rho_0$, the system undergoes a first order phase transition. While for large values of $\rho_0$, the transition disappears. 
		\textbf{(b)} The effect that the bias of the smaller subgroup has on the consensus opinion within strongly biased minority and weaker majority. When the smaller subgroups bias is significantly stronger than the larger subgroups, the consensus opinion shifts to the negative values of magnetisation. Otherwise, the majority wins. Only in a very narrow range of $\omega_-$ one can observe a first order phase transition. Otherwise, the behaviour is similar to that of the Ising model in an external field. 
	}
	\label{fig4}
\end{figure}

In figure~\ref{fig4}(a) we show how unbiased individuals affect the consensus opinion in a system with a stronger minority and weaker majority. 
For low temperatures, the system resides in a state with strong negative magnetisation, which can be 
interpreted as the stronger minority dictating the consensus. 
As the temperature increases, the system undergoes a first order phase transition from the state with strong negative magnetisation to the state with weak positive magnetisation. 
With a further increase in temperature, the system approaches the state $m=0$, although no phase transition occurs at this point.

We observe this kind of phase transition only for small values of the fraction of unbiased individuals. 
As $\rho_0$ increases, the transition disappears and the system starts to behave similarly to the Ising model in an external field. 
The jump in magnetisation, or equivalently the latent heat, decreases with an increase in the number of unbiased individuals. 
This correlates well with the aforementioned conclusion from reference~\cite{Hartnett16} that unbiased individuals help in reaching a consensus.

In figure~\ref{fig4}(b), the effect of the minority's bias on the consensus is shown.
When the bias is strong (low values of $\omega_-$) the minority dictates the opinion for all values of the temperature. 
On the other hand, if the minority is not strong enough, the resulting opinion is completely in the positive~$m$ semi-plane. 
Only for a small range of $\omega_-$, a transition occurs. 
In order to observe this kind of behaviour, two factors must be met: the fraction of unbiased individuals $\rho_0$ should be small, and the biases of two competing subgroups $\omega_+$ and $\omega_-$ must be fine-tuned.

\subsubsection{Decision making in two confronting strongly biased groups}

The third case we consider is the system where two opposing subgroups have very strong biases. 
To this end, we set $\rho_+=0.7$, $\rho_-=0.3$ and $\omega_+=10$. 
Similarly to the previous two cases, we keep an asymmetric distribution between the positively biased and negatively biased subgroups.

\begin{figure}[htbp]
	\centerline{
		\includegraphics[width=0.45\textwidth]{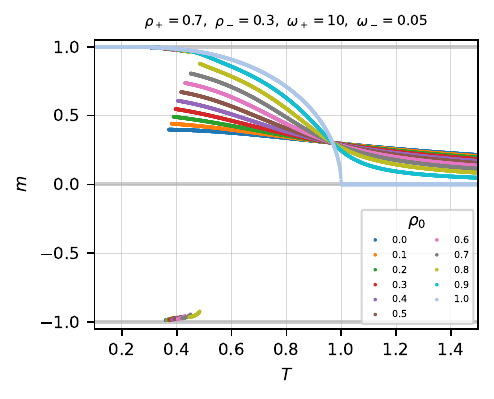}
		\includegraphics[width=0.45\textwidth]{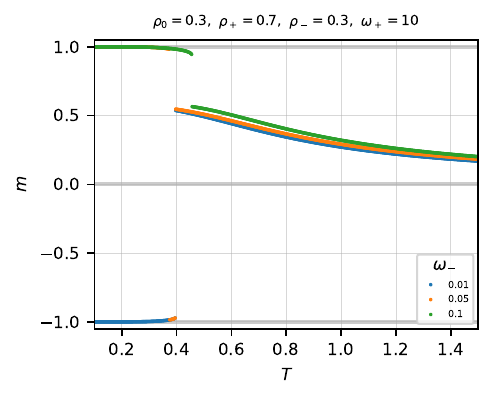}
	}
	\centerline{\textbf{(a)} \hspace{15em} \textbf{(b)}}
	\caption{(Colour online) Equilibrium magnetisation as a function of temperature in the case when the system consists of two confronting subgroups with very strong biases. In this case, for some sets of parameters the system might undergo a re-entrant phase transition. 
		\textbf{(a)} The effect that the fraction of unbiased individuals has on the consensus opinion within two very strongly biased subgroups. Similarly to the previous case shown in figure \ref{fig4}(a), an increase in the number of unbiased individuals results in a more aligned equilibrium state for the temperature range outside the transition region. For low values of $\rho_0$, the system undergoes a re-entrant phase transition when for a short range of temperatures the system jumps from a positive magnetisation semi-plane to negative values of~$m$, while for large values of $\rho_0$, the transition disappears. Another interesting aspect is the existence of a temperature for which the equilibrium opinion does not depend on~$\rho_0$. 
		\textbf{(b)} The effect that the bias of the smaller subgroup has on the consensus formation between two very strongly biased subgroups. For very low values of $\omega_-$ as well as for larger ones, only a single first order phase transition is observed, while the re-entrant phase transition occurs in a narrow range of $\omega_-$.
	}
	\label{fig5}
\end{figure}

In figure~\ref{fig5}(a) we show how the fraction of unbiased individuals changes the equilibrium magnetisation dependency on temperature. 
On the one hand, like in the previous two cases, an increase in~$\rho_0$ results in a more aligned consensus opinion. 
However, this is only the case up to a certain temperature, starting from which the trend is the opposite: an increase in~$\rho_0$ decreases the magnetisation. 
On the other hand, as temperature increases, the system undergoes two consecutive phase transitions from large positive values of~$m$ to large negative values and then back to large positive values. 
This re-entrant phase transition is a new feature inherent to this case and not observed in the previous two cases.
Similarly, as $\rho_0$ becomes large, both phase transitions disappear. 
Unfortunately, the phase with large negative magnetisation is only present for a very small temperature region, making it very hard to observe in simulations (as we will see later).

Another interesting aspect of figure~\ref{fig5}(a) is that there exists a temperature for which the equilibrium magnetisation does not depend on the fraction of unbiased individuals: the intersection of all curves in the figure.

In figure~\ref{fig5}(b) we examine the effect that $\omega_-$ has on the equilibrium magnetisation. 
For $\omega_-=[0.01,0.04]$, a single first order phase transition takes place between the state with large negative magnetisation and a state with positive magnetisation. 
Similarly, for $\omega_-=[0.07,0.1]$, there is only one first order transition, but it occurs between two states with positive magnetisation. 
In the intermediate region illustrated by $\omega_-=0.05$ and $\omega_-=0.06$ one can observe a re-entrant phase transition when, with the increase in temperature, the system jumps from a state with large positive magnetisation to the state with large negative magnetisation and as the temperature further increases it jumps back to the positive~$m$ semi-plane. 
As $\omega_-$ increases, the region where the phase with negative magnetisation can be observed decreases until it totally vanishes and the $m(T)$ dependency is characterised by a single first order transition.

{\subsection{Evolution to steady state. Numerical simulations}}

In order to investigate collective decision making dynamics as well as the final stationary states, we performed numerical simulations of the model following the general setup discussed in~\cite{Sarkanych23}. The collective decision making process of the original model~\cite{Hartnett16} is stochastic due to the probabilistic interaction mechanism: An agent aligns with its neighbours' social opinion field with the sigmoidal probability function $G_i$ given in equation~(\ref{2.2}). The nonlinearity parameter~$b$ controls the steepness of the sigmoidal function.

In particular, we expanded the original model of Hartnett {et al.}~\cite{Hartnett16} by an additional noise process to explicitly control the stochasticity, and thus the effective temperature, of the decision making process, independent of the interaction.
Via this noise process, an agent has a probability $p_\text{noise}$ of switching from state $\pm 1$ to $\mp 1$ within a time step.
For $p_\text{noise} = 0$, the original model on a complete graph is recovered.
{Even though the temperature and the noise parameter are related, this relation is rather qualitative: an increase in temperature plays the role similar to an increase in~$p_\text{noise}$.}

In individual-based model simulations, we necessarily always simulate a finite system.
Thus, besides testing the theoretical predictions in the thermodynamic limit, our simulations also allow us to study potential finite size effects, such as relaxation towards long-lived meta-stable states, resulting in a co-existence of different steady states and dependence on initial conditions~\cite{Sarkanych23}.
Therefore, in addition to the steady-state average opinion state (magnetisation~$m$), we also evaluate its standard deviation across many realisations to quantify the variability of the collective decision process due to noise and finite size effects. 

\medskip

{\subsubsection{Numerical implementation}}
The computer model used for numerical simulations was implemented in the \emph{Utopia} modelling framework~\cite{Riedel20,Sevinchan20,Sevinchan20boosting}, which handles the simulation configuration, parallelised parameter sweeps as well as efficiently reading, writing, and evaluating high-dimensional simulation output.
The model itself was implemented in modern~C++ and uses a graph data structure to allow representing arbitrary topology, while also including performance improvements for the algorithmically simpler complete graph scenario.
The model implementation can be found online, see~\cite{Sevinchan24model}.

At the beginning of each realisation, agents are grouped according to $\rho_0$ and $\rho_\pm$ and respective bias values $\omega_0 = 1$ or $\omega_\pm$ are assigned.
Each agent is randomly assigned a magnetisation $S_i$ so that the average magnetisation matches the specified $m_0 \coloneqq m(t=0)$ initialisation parameter.
During the model iteration, each agents' social field~$h_i$ is computed using equation~(\ref{2.1}).
Depending on their current state, agents will flip to $S_i=+1$ with a probability of $G_i$ [see equation~(\ref{2.2})] or to $S_i=-1$ with a probability of $1-G_i$.
Subsequently, through the additional noise process, agents have a probability of $p_\text{noise} \in [0, 1]$ (also referred to as noise level) to flip to the opposite state.
Finally, all state changes are applied synchronously and the model continues to the next iteration step.
Iteration is repeated until the specified number of iteration steps is reached, concluding one realisation.
For the next realisation, the agents' biases and states are randomly re-initialised and the above procedure is repeated.
For each parameter combination, 512~realisations were carried out.

In all simulations, we used complete graphs with $N=10^4$ agents.
If not mentioned otherwise, we used $b=1.0$ for the nonlinearity parameter; as can be seen in figure~\ref{fig2}(a), this corresponds to an almost linear mapping between $h_i$ and $G_i$.
The initial magnetisation $m_0$ for the different parameter scans performed was chosen so that relaxation towards meta-stable states, as described in \cite{Sarkanych23}, is reduced whenever possible. In particular, when the expected final steady state magnetisation across the parameters scanned is predicted by the theory to be $m>0$ ($m<0$), then we choose the initial state to be ordered with $m_0=+1$ ($m_0=-1$).
On the other hand, if the sign of the equilibrium opinion is predicted by the theory to vary across the parameters scanned, then we choose the disordered initial condition with $m_0=0$.

To numerically determine the final state magnetisation, we ran simulations sufficiently long and used values from the end of the resulting time series $m(t)$ to compute the final state magnetisation.
With the system reaching its pseudo-steady state within at most 200~iteration steps, we chose a simulation time of 1000~steps. To reduce the effect of the noise process on the quantitative result, we averaged the magnetisation~$m$ over the last 50~steps to arrive at the average final state magnetisation.

\medskip

\subsubsection{Simulation results}
In the following, we present the results of numerical simulations by showing the mean steady-state opinion and the standard deviation across realisations.

\begin{figure}[htbp]
   \centerline{
        \includegraphics[width=0.45\textwidth]{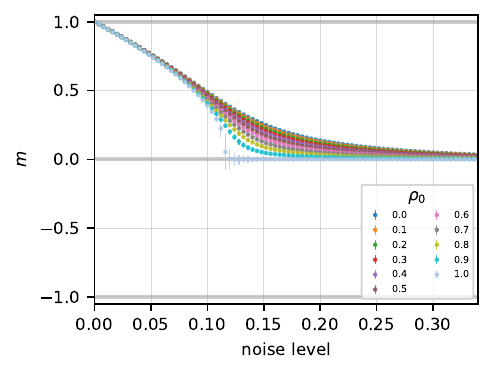}
        \includegraphics[width=0.45\textwidth]{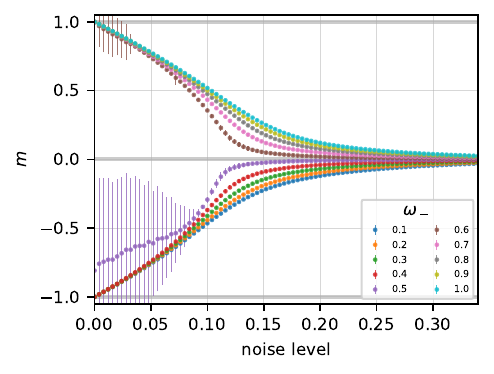}
    }
   \centerline{
        \textbf{(a)} \hspace{15em}
        \textbf{(b)}
    }
    \caption{(Colour online) 
        Two weakly biased groups; simulation results for a complete graph with $N=10^4$ nodes, showing average final state magnetisation depending on noise level.
        \textbf{(a)} Colours denote varying $\rho_0$ values, other parameters are chosen to match the analytical results in figure~\ref{fig3} ($\rho_+=0.6,~\rho_-=0.4,~\omega_+=1.5,~\omega_-=0.7$).
        The initial magnetisation is $m_0 = +1$.
        Each point is the mean value over 512~realisations, with error bars showing the corresponding standard deviation.
        \textbf{(b)} Like \emph{(a)} but for a fixed $\rho_0=0.7$ and varying $\omega_-$ parameter.
        Here, the system was initialised with $m_0 = 0$.
        The increased variability at low noise levels for $\omega_- = 0.5$ and $0.6$ illustrates the bi-stability in that parameter range.
    }
    \label{sim_weakly_biased}
\end{figure}

\begin{figure}[htb]
   \centerline{
        \includegraphics[width=0.45\textwidth]{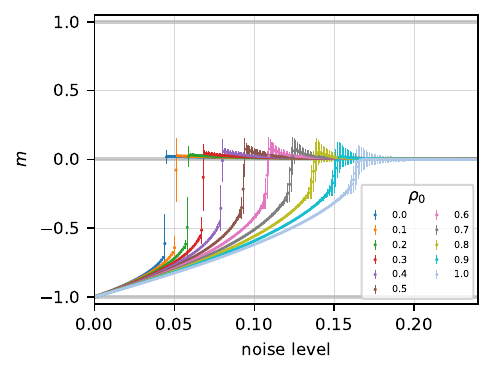}
        \includegraphics[width=0.45\textwidth]{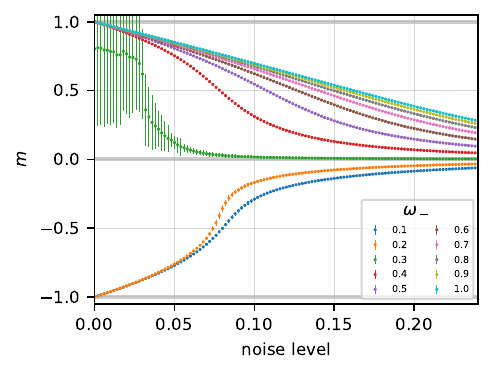}
    }
   \centerline{
        \textbf{(a)} \hspace{15em}
        \textbf{(b)}
    }
    \caption{(Colour online) 
        Strong minority vs. weak majority; simulation results for a complete graph with $N=10^4$ nodes, showing average final state magnetisation depending on noise level.
        \textbf{(a)} colours denote varying $\rho_0$, other parameters are as in figure~\ref{fig4} ($\rho_+=0.6,~\rho_-=0.4,~\omega_+=2.0,~\omega_-=0.3$).
        Here, the non-linearity parameter was set to a value ($b=1.3$) that qualitatively matches the corresponding analytical results.
        The initial state was chosen as $m_0 = -1$ to focus on the transition from negative to positive final magnetisation.
        Each point is the mean value over 512~realisations, with error bars showing the corresponding standard deviation.
        \textbf{(b)} Like \emph{(a)} but for a fixed $\rho_0=0.3$, varying $\omega_-$ instead.
        Due to the existence of both positive and negative attractive states, the initial state was chosen as $m_0 = 0$, leading to high fluctuations for the $\omega_- = 0.3$ line at low noise levels.
        Unlike in figure~\ref{fig4}(b), where $\omega_- = 0.3$ line has a discontinuous transition from $m < 0$, these simulations only show the transition from a state with $m > 0$.
    }
    \label{sim_strong_minority_weak_majority}
\end{figure}

\begin{figure}[htb]
   \centerline{
        \includegraphics[width=0.45\textwidth]{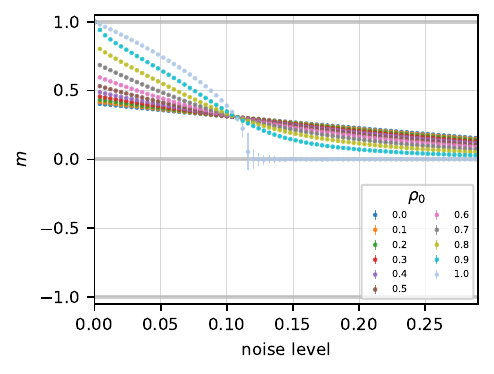}
        \includegraphics[width=0.45\textwidth]{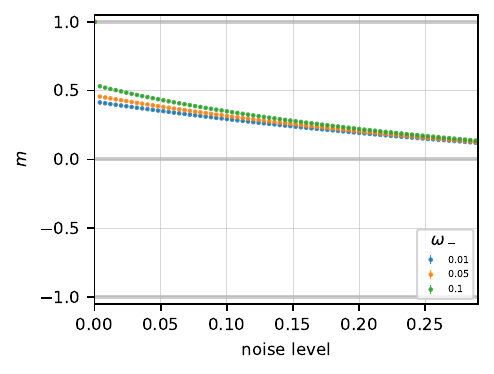}
    }
   \centerline{
        \textbf{(a)} \hspace{15em}
        \textbf{(b)}
    }
    \caption{(Colour online) 
        Two opposing groups with strong biases; simulation results for a complete graph with $N=10^4$ nodes, showing average final state magnetisation depending on noise level.
        \textbf{(a)} Colours denote varying $\rho_0$, other parameters are as in figure~\ref{fig5} ($\rho_+=0.7,~\rho_-=0.3,~\omega_+=10,~\omega_-=0.05$).
        The initial state is $m_0 = +1$.
        \textbf{(b)} Like \emph{(a)} but with $\rho_0 = 0.3$ and varying $\omega_-$.
        Note that the range $\omega_-$ is varied in is $[0.01, 0.1]$, not $[0.1, 1.0]$ as in previous figures.
    }
    \label{sim_opposing_groups}
\end{figure}

In figure~\ref{sim_weakly_biased}(a), we confirm the theoretical prediction of section~\ref{III.1}  on the impact of the ratio of unbiased agents on the dependence of magnetisation on effective temperature, here controlled via the noise intensity. Only for $\rho_0=1$, we observe a second order phase transition behaviour, with a corresponding critical point where the magnetisation drops to zero. In the presence of any biased individuals, the system exhibits a monotonous decrease of magnetisation towards $m=0$, analogous to the Ising model in an external field.

In figure~\ref{sim_weakly_biased}(b), we reproduce the theoretical results on the impact of bias strength of a minority, in the presence of an oppositely biased majority. Once the minority bias is sufficiently strong, it is able to overcome the majority, which is biased towards opinion $S=+1$, and switch the collective decision state towards negative magnetisation. Here, the most prominent difference to the theoretical prediction (figure~\ref{fig3}(b)), is the deviation of the average magnetisation for $\omega_-=0.5$ from the curves for a stronger negative bias $\omega_i<0.5$. The reason for this is the existence of a meta-stable state at positive magnetisation for finite-sized systems and the relaxation of individual simulation runs starting from a disordered initial condition $m_0=0$ towards this meta-stable state with $m>0$. Thus, due to the bistability of the steady-state solutions in numerical simulations, the average over all realisations is closer to the disordered state $m=0$. A signature of this bistability is also the large standard deviations of the average magnetisation for the corresponding parameters.  

In figure~\ref{sim_strong_minority_weak_majority}(a), we show the role of the ratio of unbiased agents for the average magnetisation in the case of a competition of a strongly biased minority to a weakly biased majority. Again our numerical simulations confirm the theoretical prediction of a discontinuous transition, with the discontinuity decreasing with increasing $\rho_0$, eventually resulting in a continuous transition for $\rho_0 \to 1$.

In figure~\ref{sim_strong_minority_weak_majority}(b), the competition of the oppositely biased, differently-sized subgroups is investigated for different bias strengths $\omega_-$, of the negatively biased minority ($\rho_-=0.4$ versus $\rho_+=0.6$).
As predicted by the theory, for a sufficiently strong negative bias ($\omega_-\lesssim 0.3$), the minority is able to dominate the collective opinion. The main difference to the analytical prediction [figure~\ref{fig4}(b)] is the positive overall magnetisation for $\omega_-=0.3$. This is probably a consequence of the initial condition and the finite size of the system. For this transitional negative bias, the state with the positive consensus opinion appears to be a meta-stable state with a large basin of attraction. For the chosen initial condition $m_0=0$, a majority of individual simulation runs relax towards this positive meta-stable state; only a minority of runs relax towards a negative steady state as predicted by the theory, which results in a  large standard deviation of the mean magnetisation when averaged over multiple realisations.

In figure~\ref{sim_opposing_groups}(a), we show that our numerical simulations are capable of reproducing the theoretical result on all steady-state opinion solutions for different $\rho_0$, to cross each other at a particular value of effective temperature, controlled here by the noise level. Below the crossing point ($p_\textnormal{noise}\lesssim 0.1$), an increase in the ratio of unbiased individuals leads to an increase in average opinion, while above the crossing point ($p_\textnormal{noise}\gtrsim 0.1$) the average consensus opinion decreases with $\rho_0$.

In figure~\ref{sim_opposing_groups}(b) we investigate the competition of two strongly biased groups for different bias strengths of the negatively biased minority. In contrast to the theoretical predictions, for the system size studied, we do not observe the discontinuous transition or the re-entrant phase transition at low temperatures. The consensus opinion is dominated by the positively biased majority with a monotonous decrease of $m$ with an increasing noise level.
This is probably due to finite size effects, which become particularly impactful for low noise, where the phase transitions had been predicted. An additional difficulty is the unknown mapping between the temperature in the Hamiltonian and the noise level~$p_\textnormal{noise}$ in the simulations. Even for very large systems, an extremely fine sampling at very low~$p_\textnormal{noise}$ may be required to observe the predicted effects.

\newpage
\section{Conclusions and outlook}\label{IV}

Recently \cite{Sarkanych23}, we have extended the model suggested by Hartnett~{et~al.}~\cite{Hartnett16}
to describe consensus formation. The original model rests on two ingredients: agent inhomogeneity and non-linearity of their interaction.
In our extension, in addition to the randomness in the agent bias which is described in the original model by random variables $\omega_i$, we have accounted for yet another random feature: fluctuations in agent opinion, a noise process that can randomly change the agent opinion irrespective of the opinions of its neighbours.
In our equilibrium statistical physics description, we attribute this feature to the temperature~$T$.
As we show in this paper, even if considered on a complete graph, the model manifests rich and sometimes unexpected behaviour for non-zero temperature. 
The last, in turn, may be interpreted as new emergent critical behaviour induced by noise in a multi-agent system. Having in mind all possible caveats of such an analogy, we support our observations by a series of numerical simulations, where the noise was directly incorporated on each step of the algorithm in a probabilistic way.
Besides the reported novel phase transitions, a specific result we would like to highlight here, is the different impact of uninformed agents on the steady state opinion for the strongly-biased groups scenario: For low temperatures, the steady state opinion increases with $\rho_0$, but above a specific temperature where the consensus state does not depend on $\rho_0$, the opposite is the case.
The qualitative and very often close quantitative correspondence between analytic predictions and results of numerical simulations is encouraging and calls for further analysis. 
In particular, an obvious further step will be to use the model to analyse consensus decision making on graphs with a more complex topology. 

{\section*{Acknowledgements}}

We write this paper for the special issue of \textit{Condensed Matter Physics} in memory of Ralph Kenna.
His academic activities covered a broad field of statistical physics and complex system science, in
particular sociophysics and cultural complexity where he developed quantitative methods in analysing the mythological
narratives  \cite{Kenna12,Kenna17,Kenna22}, for which he was aptly called a Father of mythematics. In this sense our paper where
we use statistical physics to understand biological systems is somewhere in between the Ralph’s
interests. One of the results of his devoted work was the $\mathbb{L}^4$ collaboration \cite{L4}, where some of the
authors are affiliated. In this sense this our work continues what he had initiated.
We will miss Ralph in physics, in mythematics, in $\mathbb{L}^4$, and not only. 
We deeply acknowledge what he has done and thank him for that.

The specific research presented here, was supported by the BMBF Bridge2ERA program, Project 01DK20044 ('Complex networks: self-organization and collective information processing'),  Deutsche Forschungsgemeinschaft (DFG, German Research Foundation) under Germany's Excellence Strategy-EXC 2002/1 'Science of Intelligence', Project 390523135 (Y.S. and P.R.), and National Research Foundation of Ukraine, Project 246/0099 “Criticality
of complex systems: fundamental aspects and applications” (P.S., M.K., Yu.H).
Furthermore, we acknowledge support by Andrew Hartnett for clarifications on his model and simulation details.

\bibliographystyle{cmpj}
\bibliography{cmpjxampl} 

\newpage

\ukrainianpart

\title{Консенсусне прийняття рішень на повному графі: складна поведінка на основі простих припущень}
\author{{П. Сарканич}\refaddr{inst1,inst2}, {Ю. Севінчан}\refaddr{inst3,inst4}, {М. Красницька}\refaddr{inst1,inst2}, 
{П. Романчук}\refaddr{inst3,inst4}, {Ю.~Головач}\refaddr{inst1,inst2,inst5,inst6}}
\addresses{
	\addr{l1}{Інститут фізики конденсованих систем НАН України, Львів, 79011, Україна}
\addr{l2}{Співпраця $\mathbb{L}^4$ і Коледж докторантів ``Статистична фізика складних систем'', Ляйпціґ-Лотаринґія-Львів-Ковентрі, Європа}
\addr{l3}{Кафедра біології, Інститут теоретичної біології, Берлінський університет імені Гумбольдтів, Берлін, 10099, Німеччина}
\addr{l4}{Дослідницький кластер передових досліджень ``Наука про інтелект'', Берлін, 10587, Німеччина}
	\addr{inst7}{Haiqu, Inc., вул. Шевченка 120Г, Львів, 79039, Україна}
 	\addr{l5}{Центр плинних і складних систем, Університет Ковентрі, Ковентрі, CV1 5FB, Велика Британія}
    \addr{l6}{Центр науки про складність, Відень, 1080, Австрія}
}

%
%
%

\makeukrtitle

\begin{abstract} 
У статті ми досліджуємо модель одностайного (консенсусного) прийняття рішень
[Hartnett A. T., et al., {Phys. Rev. Lett.}, 2016, \textbf{116}, 038701]
слідуючи підходу статистичної фізики представленому в [Sarkanych P.,
et al., {Phys. Biol.}, 2023, \textbf{20}, 045005]. У такому підході 
температура є мірою флуктуацій, які не розглядалися раніше  в оригінальній моделі. Ми розглядаємо модель на повному 
графі. Основна мета статті полягає в тому, щоб показати, що аналітичний опис може призвести до дуже багатої фазової 
поведінки, неочікуваної для моделей на  повному графі. Однак різноманіття індивідуальних ознак агента (спіну) --- неоднорідність і сила переконання --- враховані моделлю призводять до  нетривіальних колективних ефектів. Ми демонструємо, що ці ефекти можуть виникати як  неперервні або різкі фазові переходи, які іноді супроводжуються повторенням (re-entrant behaviour) чи перевертанням параметрів порядку. Це у свою чергу може призвести до цікавої інтерпретації в термінах прийняття соціальних рішень. Ми підтверджуємо аналітичні прогнози за допомогою чисельного моделювання. В той час як аналітичні розрахунки виконуються в рамках формалізму рівноважної статистичної фізики, чисельне моделювання додає ще одну динамічну особливість --- локальну нелінійність або підпорядкування індивіда думці його оточення. Схоже, що ця особливість має сильний вплив як на спосіб досягнення рівноважного стану, так і на його характеристики.

\keywords 
			колективне прийняття рішень, спінові моделі, уподобання, узгодження 
\end{abstract}

\end{document}